\documentclass[nofootinbib,amsfonts,prd,aps]{revtex4}
\usepackage{graphicx}
\begin{document}

\title{Hawking radiation from a spherical loop quantum gravity black hole}

\author{Rodolfo Gambini$^{1}$,
Jorge Pullin$^{2}$}
\affiliation {
1. Instituto de F\'{\i}sica, Facultad de Ciencias, 
Igu\'a 4225, esq. Mataojo, 11400 Montevideo, Uruguay. \\
2. Department of Physics and Astronomy, Louisiana State University,
Baton Rouge, LA 70803-4001}

\begin{abstract}
  We introduce quantum field theory on quantum space-times techniques
  to characterize the quantum vacua as a first step towards studying
  black hole evaporation in spherical symmetry in loop quantum gravity
  and compute the Hawking radiation. We use as quantum space time the
  recently introduced exact solution of the quantum Einstein equations
  in vacuum with spherical symmetry and consider a spherically
  symmetric test scalar field propagating on it. The use of loop
  quantum gravity techniques in the background space-time naturally
  regularizes the matter content, solving one of the main obstacles to
  back reaction calculations in more traditional treatments. The
  discreteness of area leads to modifications of the quantum vacua,
  eliminating the trans-Planckian modes close to the horizon, which in
  turn eliminates all singularities from physical quantities, like the
  expectation value of the stress energy tensor. Apart from this, the
  Boulware, Hartle--Hawking and Unruh vacua differ little from the
  treatment on a classical space-time. The asymptotic modes near scri
  are reproduced very well. We show that the Hawking radiation can be
  computed, leading to an expression similar to the conventional one
  but with a high frequency cutoff. Since many of the conclusions
  concern asymptotic behavior, where the spherical mode of the field
  behaves in a similar way as higher multipole modes do, the results
  can be readily generalized to non spherically symmetric fields.
\end{abstract}

\maketitle

\section{Introduction}

The evaporation of a black hole is one of the most fascinating
problems in fundamental theoretical physics today, as can be attested
by the surge of activity related to ``firewalls'' \cite{amps} in the
last few months. A complete treatment of the evaporation requires a
theory of quantum gravity. Loop quantum gravity is a contender for
such a theory, but a complete treatment of the evaporation has proved
elusive. Here we take an incremental step in its study by considering
a quantum field theory on quantum space-time approach \cite{akl,dlt},
studying a spherically symmetric scalar field propagating on the
recently introduced exact solution for the quantum space-time of a
vacuum spherically symmetric black hole\cite{sphericalprl}.  We will
treat the matter field as a test field, as a first step towards a
perturbative treatment of the evaporation via back-reaction.  We
consider the quantum states to be a direct product of states of
gravity and states of matter. For the states of gravity we take the
physical states constructed in \cite{sphericalprl} that are
annihilated by the Hamiltonian and diffeomorphism constraints of
vacuum spherically symmetric gravity.  We take the expectation value
of the matter part of the Hamiltonian constraint on the exact quantum
states of the gravitational field, and write it in terms of
parameterized Dirac observables of the gravitational field. The
resulting operator acts on the states of matter as a true Hamiltonian
yielding quantum gravity corrected equations for the propagation of
matter. The main quantum gravity correction consists in the
discretization of the equations of motion as a consequence of the
discrete structure of space in loop quantum gravity. We study the
impact of the resulting changes on the various usual vacua for quantum
fields in a Schwarzschild space-time. All of them suffer small
modifications due to the discreteness. Also, all issues involving
singularities of physical quantities at horizons are resolved by the
discreteness. We study the Hawking radiation in terms of two point
functions taking into account the discreteness induced by the
quantization and show that the black body radiation is a robust
property.

The organization of this paper is as follows. In section II we review
the recent developments concerning the solution of vacuum spherically
symmetric loop quantum gravity and its corresponding space of physical
states. In section III we review the use of parameterized Dirac
observables to represent variables in totally constrained systems and
how they are applied in this case. In section IV discuss the
Hamiltonian for the matter fields and how to realize it as a quantum
operator on the space of physical states discussed in section
II. Section V discusses the resulting equations of motion corrected
due to the quantum background space-time. Section VI proceeds to
discuss the various quantum vacua that are usually considered in the
context of black holes in light of the corrected evolution
equations. Section VII discusses the Hawking radiation.  We end in
section VIII with conclusions.

\section{Spherically symmetric vacuum gravity}

We briefly review spherically symmetric vacuum gravity, referring the
reader to our previous work \cite{sphericalprl} for further  references. 
The Ashtekar-like variables adapted to spherical symmetry yield
two pairs of canonical variables $E^\varphi$,
${K}_\varphi$ and $E^x$, $K_x$, that are related to the traditional
canonical variables in spherical symmetry $ds^2=\Lambda^2 dx^2+R^2
d\Omega^2$ by $\Lambda=E^\varphi/\sqrt{|E^x|}$, $P_\Lambda=
-\sqrt{|E^x|}K_\varphi$, $R=\sqrt{|E^x|}$ and $P_R=-2\sqrt{|E^x|} K_x
-E^\varphi K_\varphi/\sqrt{|E^x|}$ where $P_\Lambda, P_R$ are the
momenta canonically conjugate to $\Lambda$ and $R$ respectively, $x$
is the radial coordinate and $d\Omega^2=d\theta^2+\sin^2\theta
d\varphi^2$. We will take the Immirzi parameter equal to one.

As discussed in \cite{sphericalprl}, rescaling and combining the
Hamiltonian and diffeomorphism constraints leads to the following
total Hamiltonian,
\begin{eqnarray}
H_T &=&\int dx \left[ -N'
\left(-\sqrt{E^x}\left(1+K_\varphi^2\right)+\frac{\left(\left(E^x\right)'\right)^2\sqrt{E^x}}{4
    \left(E^\varphi\right)^2}+2 G M\right)\right.\nonumber\\
&&
+ N \left(\frac{1}{8} \frac{\left(E^x\right)'
      P_\phi^2}{\left(E^\varphi\right)^2\sqrt{E^x}}+
    \frac{\left(E^x\right)'\left(E^x\right)^{3/2}\left(\phi'\right)^2}{2\left(E^\varphi\right)^2}-\frac{K_\varphi
      \sqrt{E^x}P_\phi \phi'}{E^\varphi}\right)
+ N_r \left[-\left(E^x\right)' K_x +E^\varphi
  K_\varphi'+P_\phi \phi'\right]
\end{eqnarray}
where $N$ and $N_r$ are combinations of the original lapse and shift,
i.e., 
\begin{eqnarray}
  N^{\rm orig}&=& \frac{N \left(E^x\right)'}{E^\varphi},\\
  N_r^{\rm orig} &=& N_r -\frac{2
    K_\varphi\sqrt{E^x}}{\left(E^x\right)'}N^{\rm orig},
\end{eqnarray}
and we have added a scalar field characterized by the canonical pair
of variables $\phi, P_\phi$. 

The constraints introduced above close a Lie algebra, and this allows
to complete the Dirac quantization in closed form.  For the quantum
treatment of the vacuum theory we refer the reader to
\cite{sphericalprl}. We would like to consider a state that
approximates well a classical metric with a given value of the mass
$M_0$, it will be given by a superposition
\begin{equation}
  \vert \psi,\tilde{g},\vec{k}\rangle_{\rm grav} =\int
  \vert\tilde{g},\vec{k},M\rangle\psi(M) dM
\end{equation}
with $\psi(M)= c
\exp\left(\frac{-\left(M-M_0\right)^2}{2\sigma}\right)$ with $\sigma$
small compared to $M_0^2$ and $c$ a normalization constant. As a first
step, we will assume that the spread of the mass is negligible and all
expressions will be analyzed only at leading order in it, that is, we
neglect fluctuations in the mass. The state $\vert
\tilde{g},\vec{k},M\rangle$ is an exact solution of the constraints as
constructed in \cite{sphericalprl}. On such states $M$, a Dirac
observable, acts as an operator multiplicatively. We recall that
$\tilde{g}$ is a family of graphs related by diffeomorphisms and
$\vec{k}$ are the valences of the links in the spin nets. The
gravitational part of the Hamiltonian vanishes exactly on such a
state. We will discuss the choice of $\vec{k}$ later on which achieves
in the simplest terms a semiclassical behavior in the exterior and
near the horizon. We are choosing a state with well defined values of
$\vec{k}$, one could have considered superpositions. Either
considering superpositions of the mass or the $\vec{k}$'s does not
change the results discussed here provided one is superposing states
that approximate a classical geometry well. It should be noted that
obtaining these physical states does not involve any gauge
fixing (apart from using coordinates adapted to spherical symmetry).

Our general philosophy is to treat the scalar field as a test field in
a background defined by a semiclassical state that approximates the
geometry of a Schwarzschild black hole in vacuum. To this aim, we will
assume that the quantum states are a tensor product of states of
vacuum gravity and states of matter.  We will evaluate the expectation value
of the matter part of the Hamiltonian on a state of vacuum
gravity. The resulting quantity is an operator acting on the matter
variables and we will interpret it as the quantum Hamiltonian of
matter living on a quantum space-time.

\section{Parameterized Dirac observables}

We need to write the matter part of the Hamiltonian as a Dirac
observable to compute its expectation value in the physical space of
states corresponding to a black hole in vacuum. That would make it a
well defined quantity to promote to an operator on the space of
physical states of the vacuum gravity theory and it would make it
commute with the pure gravity part of the constraint. To this aim, we
will use the technique that Rovelli \cite{rovelli} calls ``evolving
constants of motion'', which can perhaps be better characterized in
this context as ``parameterized Dirac observables'' \cite{4models}.
It was originally developed to address the time evolution of
constrained mechanical systems but can be extended to any constrained
field theory. In that case it can be used to discuss local properties
of dynamical variables that are gauge dependent.  When one is dealing
with constrained field theories, the dynamical variables of the theory
are generically not defined as operators acting on the physical space
of states. However, they can be written as functions of the Dirac
observables and some (functional) parameters and then can be viewed as
acting on the physical space of states. This in particular applies in
the gravitational case. Although there is a tendency to believe that
the only well defined quantum operators in the gravitational case, due
to diffeomorphism invariance, will be global quantities, it is
possible to describe local properties of the space-time as a function
of the (global) observables and parameters.

Let us recall the definition of parameterized Dirac observables. They
are functions of the canonical variables and parameters that have
vanishing Poisson brackets with all the constraints of the system.  We
illustrate the technique with a simple mechanical example: the
parameterized free particle. One has canonical variables
$x_0,x_1,p_0,p_1$ and a constraint $C=p_0+p_1^2/(2m)=0$. A pair of
independent Dirac observables are $X=x_1-x_0 p_1/m$ and $p_1$
(or $p_0$). The physical states, annihilated by the constraints are
given by $\psi(p_0,p_1)=f(p_1)\delta(p_0+p_1^2/(2m))$ in the momentum
representation, with $f(p_1)$ an arbitrary function. The inner product
on the phase of physical states obtained via refined algebraic
quantization is given by 
\begin{equation}
  \langle \psi_1 \vert \psi_2\rangle= \int_{-\infty}^{\infty} dp_0
  \int_{-\infty}^\infty dp_1 \delta(p_0+p_1^2/(2m))f_1^*(p_1)f_2(p_1)=
\int_{-\infty}^\infty dp_1 f_1^*(p_1)f_2(p_1),
\end{equation}
yielding the ordinary inner product of the non-relativistic particle in
quantum mechanics.  A dynamical variable like $x_1$ cannot be directly
represented on the space of physical states. However, one can
construct a parameterized Dirac observable
$X_1(\lambda)=X+(p_1/m)\lambda$ with $\lambda$ a parameter. This has the
property that $X_1(\lambda=x_0)=x_1$, and in that sense we consider
$X_1$ to be a representation of $x_1$. Notice that the quantization
obtained could have been derived using a gauge fixing $x_0=\tau$,
which imposed as a constraint leads to the fixing of the Lagrange
multiplier and one would obtain in the Heisenberg representation for
$x_1$ the same quantum representation as the one obtained here, with a
particular value of the parameter $\lambda=\tau$. It is important to
emphasize that the parameterized Dirac observables must be self
adjoint operators. For instance, if one had tried to build a
parameterized Dirac observable $X_0(\lambda)$ one would find it is not
self-adjoint. The condition of self adjointness in this case leads to
the identification of a good time variable for the system. Further
details can be seen in reference \cite{4models}.

At a classical level, in vacuum spherically symmetric gravity there
are two constraints and two (functional) parameters, with $M$ the
Dirac observable. We can define a parameterized Dirac observable
associated with $E^\varphi$ guided by the expression of the
Hamiltonian constraint,
\begin{equation}
  E^\varphi({\cal K}_\varphi,{\cal E}^x)= 
\frac{\left({\cal E}^x\right)'}{2\sqrt{1+{\cal
      K}_\varphi^2-\frac{2GM}{\sqrt{{\cal E}^x}}}},
\end{equation}
and taking ${\cal K}_\varphi$ and ${\cal E}^x$ as parameters (we write
them in calligraphic to emphasize that point, they are not canonical
variables anymore). One has, using the expression of the mass function
in terms of the canonical variables given by the Hamiltonian
constraint, that 
$E^\varphi({\cal K}_\varphi=K_\varphi, {\cal
  E}^x=E^x)=E^\varphi(x,t)$. 

Similarly, guided by the expression of the diffeomorphism constraint,
one can write a parameterized Dirac observable associated with $K_x$, 
\begin{equation}
  K_x({\cal K}_\varphi,{\cal E}^x)= \frac{\cal{K}_\varphi'}{2
    \sqrt{1+{\cal K}_\varphi^2-\frac{2GM}{\sqrt{{\cal E}^x}}}},
\end{equation}
with $K_x({\cal K}_\varphi=K_\varphi,{\cal E}^x=E^x)=K_x(x,t)$ using
the explicit expression of the observable $M$ in terms of the
dynamical variables.

As we discussed in the example of the free particle, if one wishes to
obtain the Lagrange multipliers, one can impose as constraints
$K_\varphi={\cal K}_\varphi(x,t)$ and $E^x={\cal E}^x(x,t)$ with
${\cal K}_\varphi$ and ${\cal E}^x$ given functions. The consistency
in time of these constraints will determine the Lagrange
multipliers. In particular if one chooses the given functions to be
time independent one gets that the shift vanishes and the Lapse is a
constant. It should be emphasized that there are no further
consistency conditions that need to be checked, given the fact that
the parameterized Dirac observables are written in terms of Dirac
observables. 

The quantization is a bit more delicate in this case than in the
simple example we considered before, since Dirac observables arise at
the quantum level that do not have a classical counterpart. This in
particular will imply that the parameterized Dirac observables are not
functions of ${\cal K}_\varphi$ and ${\cal E}^x$ but of slightly
different variables. Let us recall that on the space of states
annihilated by the Hamiltonian constraint of vacuum gravity (but not
necessarily by the diffeomorphism constraint) $\hat{E}^x$ is a well
defined operator,
\begin{equation}
\hat{E}^x(x) \vert g,\vec{k},M\rangle =   \ell_{\rm Planck}^2 k(x) \vert g,\vec{k},M\rangle, 
\end{equation}
where $k(x)$ is the value of the valence of the spin network between
the two consecutive vertices of the spin network that contain within
them the point $x$. To solve the diffeomorphism constraint, one group
averages. Although on the space of solutions of all the constraints
$k(x)$ is not well defined, the succession of values $\vec{k}$
is. This allows us to introduce the observable $ \hat{O}(z)\vert
\vec{k},\tilde{g}\rangle_{\rm phys}= \ell_{\rm Planck}^2 k_{{\rm
    Int}(V z)} \vert \vec{k},\tilde{g}\rangle_{\rm phys}$ with $z\in
[0,1]$ and where $V$ is the number of vertices in the spin network.
${\rm Int}(V z)$ is the integer part of the number of vertices times
$z$.  This observable allows us to encode the information in $E^x$ in
terms of it and a functional parameter $z(x)$  (a function from
the real line into the interval $[0,1]$) as $\hat{E^x}(x) =
\hat{O}(z(x))$. With this definition, $E^x$ becomes a Dirac
observable. For a choice of $z(x)$ the action of $\hat{E}^x$ on a
state of the physical space $\vert \tilde{g},\vec{k},M\rangle$ is the
same as the action of $E^x$ on the states that only solve the
Hamiltonian constraint $\vert g,\vec{k},M\rangle$. To put it
differently, the freedom present in $z(x)$ corresponds to the freedom
of spatial diffeomorphisms and to choose a $z(x)$ corresponds to a
choice of coordinates.

So the operators associated with the parameterized Dirac observables,
acting on the space of physical states are,
\begin{eqnarray}
  \hat{E}^x({\cal K}_\varphi,z(x))&=&
  \frac{\hat{O}(z(x))'}{2\sqrt{1+{\cal
        K}_\varphi^2-\frac{2G\hat{M}}{\sqrt{\hat{O}(z(x))}}}},\\
\hat{K}_x({\cal K}_\varphi,z(x))&=&
\frac{{\cal K}_\varphi'}{2\sqrt{1+{\cal
        K}_\varphi^2-\frac{2G\hat{M}}{\sqrt{\hat{O}(z(x))}}}}.
\end{eqnarray}

With this, we can rewrite the coefficients of the matter Hamiltonian
dependent on the gravitational variables as parameterized Dirac
observables acting on the space of physical states as functions of the
functional parameters ${\cal K}_\varphi,z(x)$ and the observables $M$
and $O(z)$.  We have therefore succeeded in expressing the matter
Hamiltonian entirely in terms of Dirac observables of the vacuum
gravitational theory and (functional) parameters. The functional
freedoms represented by $z(x)$ and ${\cal K}_\varphi$ correspond to
the freedoms associated with spatial diffeomorphisms and the foliation
choice. Choosing these functions is equivalent to fixing a gauge and
no further conditions are needed since everything is already expressed
in terms of Dirac observables. This is a main advantage of working
with parameterized Dirac observables, one works directly on the space
of physical states and gauges can be fixed easily, without additional
consistency conditions.

From now on we will assume (except for a brief discussion in section
V.E) that ${\cal K}_\varphi$ is time independent, further partially
fixing the gauge, again in order to ultimately deal with coordinates
in which the background is manifestly static.

\section{The Hamiltonian for the scalar field}

\subsection{A choice of the background quantum state}
We need to make a choice of the background quantum space-time. We wish
to have a quantum state that approximates a semiclassical geometry
well. We also make some additional assumptions to simplify
calculations. For instance, one could consider states that involve
superpositions of $\vec{k}$'s but we will choose not to do so. Since
all the parameterized Dirac observables only depend on $\hat{O(z(x))}$
and not its canonically conjugate variable, states with well defined
$\vec{k}$'s are eigenstates of the observables.

We also need to make a choice of the labels of the spin network
$\vec{k}$. Let us recall that the kinematically, $k_i$'s are the
eigenvalues of the $\hat{E^x}$ operator and that classically
$E^x=r^2$, that is, it is proportional to the area of the spheres of
symmetry. Therefore if one is to approximate a semiclassical
space-time one should choose $k_i's$ that monotonically grow and such
that the ``step'' between successive $k_i$'s is small compared to
their values. Obviously, there are many possible choices of $\vec{k}$
that accomplish this, and they will correspond to semiclassical states
with relatively similar properties, at least when probed at large
scales compared to the spacing of the $k_i$'s.

Whatever the choice of $\vec{k}$'s, they are constrained by the fact
that in loop quantum gravity areas are quantized. If the radial
coordinate is $r$, the difference in values for it in two successive
points of the lattice is bounded below by $\ell_{\rm
  Planck}^2/(2r)$. So in the exterior of the black hole we can choose,
for instance, a uniform spacing with separation $\Delta> \ell_{\rm
  Planck}^2/(4 G M)$ in the $r$ coordinate, with $r$ the usual
Schwarzschild radial coordinate. Notice that for
macroscopic black holes this is much smaller than Planck's
length. This immediately will limit the existence of some trans-Planckian
modes when computing the vacua.  This is in line with what has been
observed by many authors in models, for instance in analogue black
holes \cite{visser}. In order to have a semiclassical
spin network one could also choose
to have a distribution for the values of $\vec{k}$ such that
the separation in the $r$ variable between successive vertices will be
given by a fixed value $\Delta$.

We will work with a spin network with a finite number of vertices
$V$. For simplicity, we will choose the function $z(x)$ such that the
coordinate $x$ is identified with the coordinate $r$. The spin
network vertex $i$ will therefore be located at
\begin{equation}
  x_i=(i+i_H)\Delta 
\end{equation}
with $i_H={\rm Int}(2GM/\Delta)$, so the point $i=0$ is the closest
vertex in the spin network to the horizon.  We also choose $\Delta$ in
such a way that it is a multiple of $\ell_{\rm Planck}^2$. With such
choices, the eigenvalues of $\hat{E^x}$ with the choice of $z(x)$ we
made are $k_i=x_i^2/\ell_{\rm Planck}^2$, that is $\hat{E}^x_i \vert
\psi,\tilde{g},\vec{k}\rangle_{\rm grav} =\ell_{\rm Planck}^2 k_i
\vert \psi,\tilde{g},\vec{k}\rangle_{\rm grav} =x_i^2 \vert
\psi,\tilde{g},\vec{k}\rangle_{\rm grav} $. These choices for the
gauge imply $z=x/(V\Delta)$, so $z_i\equiv z(x_i)=(i+i_H)/V$.

One can penetrate into the black hole by considering negative values
of $i$ and also allowing a non-vanishing extrinsic curvature
$K_\varphi$ from the horizon inwards, since that corresponds to a
coordinate choice that makes the horizon non-singular.  The analysis
will be simpler if $2GM=i_H \Delta$, that is, we are putting a vertex
of the spin network at the horizon. That is not a generic situation,
it just simplifies calculations. Alternatively, one can consider that
the point is not on the horizon but the separation is negligible
compared to the separation to the next vertex on the spin net and then
the results will very approximately hold.

We will see that the discreteness has implications for the types of
vacua one gets. This is a priori surprising since Hawking radiation is
a phenomenon usually associated entirely with the exterior of the
black hole, where for non Planck-sized black holes, one expects
quantum gravity effects to be negligible.

It should be emphasized that the specific spin network chosen is 
mostly in order to simplify the calculations. Many other spin networks, more
refined in the separation of the vertices, could be chosen
approximating even better the classical background. However, as we
discussed, certain degree of discreteness will always be present due
to the quantization of $\vec{k}$ and this will always have
implications near the horizon and in particular, for the behavior of
the vacua at the horizons.

\subsection{The matter Hamiltonian as a parameterized Dirac observable}

On the quantum states considered $\vert
\psi,\tilde{g},\vec{k}\rangle_{\rm grav}$, the operator associated
with $E^x_i$ is,
\begin{equation}
  \hat{E}^x_i \vert
\psi,\tilde{g},\vec{k}\rangle_{\rm grav} = \ell_{\rm Planck}^2 k_i 
\vert
\psi,\tilde{g},\vec{k}\rangle_{\rm grav},
\end{equation}
(recall that we made a specific choice for the function $z(x)$).

To realize the Hamiltonian as a quantum operator we need to realize
the operator $\hat{\left(E^x\right)'}$ and its inverse. The
realization of spatial derivatives of operators in loop quantum
gravity has been considered only a few times before, and never for a
momentum operator like $\hat{E^x}$. When we analyzed the vacuum case,
only $\hat{\left(E^x\right)'}$ appeared and we represented it by a
finite difference. Because the Hamiltonian is a scalar given by an
integral of a density, the final expression is independent of the
spacing used in the finite difference. In this case we have a further
challenge, since we will need to differentiate $\phi$,
which only exists in the vertices of the spin net. This leads us to
propose that we realize all derivatives as a finite difference between
nearest vertices of the spin network. Again, because the Hamiltonian
is the integral of a density, the actual spacing drops off from the
calculation at the end of the day. It should be noted that this is at
variance, for instance, with the spirit of the realization of the
Hamiltonian constraint proposed for the three dimensional case, where
the resulting operator does not connect neighboring vertices
\cite{thiemannqsd}, but adds extraordinary vertices adjacent to the
vertex it is acting on.  This is more in the spirit of ``Algebraic
Quantum Gravity'' \cite{thiemanngiesel}.  We have conducted extensive
tests to see that our proposal leads to excellent agreement with the
classical theory in the semiclassical regime for suitable spin
networks.  We define $\hat{\left(E^x\right)'} \equiv
\hat{\left(E^x\right)'}_i/\Delta x$, with
\begin{equation}
  \hat{\left(E^x\right)'}_i \vert \psi,\tilde{g},\vec{k}\rangle_{\rm grav} = 
\ell_{\rm Planck}^2 \left[k_{i+1}-k_{i}\right]\vert \psi,\tilde{g},\vec{k}\rangle_{\rm
  grav}.
\label{exprime}
\end{equation}
And for
$\widehat{1/\sqrt{\left(E^x\right)'}}\equiv\widehat{\sqrt{\Delta x}
/\sqrt{\left(E^x\right)'_i}}$ we have, following similar steps to
those used in loop quantum cosmology \cite{asbole}, one rewrites the
term on the left as a Poisson bracket classically and then promotes 
the Poisson bracket to a quantum commutator to yield,
  \begin{equation}
    \widehat{\frac{{\rm sgn}\left(\left(E^x\right)'\right)_i}
{\sqrt{\left(E^x\right)'_i}}} \vert \psi,\tilde{g},\vec{k}\rangle_{\rm grav}=
\frac{1}{{\ell_{\rm Planck}}}\left[\sqrt{\vert\Delta k_i+1\vert}-\sqrt{\vert\Delta k_{i}-1\vert}\right]\vert \psi,\tilde{g},\vec{k}\rangle_{\rm grav},
  \end{equation}
where $\Delta k_i\equiv k_{i+1}-k_i$. Notice that this particular
representation of the operator will avoid introducing singularities in
the matter part of the Hamiltonian in the interior of the black
hole. In this paper we are concentrating on the exterior and there one
can simply talk of $1/(E^x)'$ as an operator directly, both
definitions yield extremely close results in that region.
With this, the Hamiltonian is a sum of contributions at the vertices
$H=\sum_i H_i$, with, 
\begin{equation}
  \hat{H}_i= \hat{A}_i P_{\phi,i}^2+\hat{B}_i  
\left(\phi_{i+1}-\phi_i\right)^2+\hat{C}_i
P_{\phi,i}\left(\phi_{i+1}-\phi_i\right)
\end{equation}
where $P_{\phi,i}$ is the value of the momentum of the scalar field at
the vertex $i$ and similarly for $\phi_i$, and these two quantities
are operators acting on the quantum states of matter. The coefficients are 
\begin{eqnarray}
  \hat{A}_i \vert \psi,\tilde{g},\vec{k}\rangle_{\rm grav} &=& \frac{1}{2}
  \frac{\left[\sqrt{\vert\Delta k_i+1\vert}-\sqrt{\vert\Delta k_{i}-1\vert}\right]^2}{\ell_{\rm Planck}^2}
    \frac{\left(1+{\cal K}^2_{\varphi,i}
-\frac{2 G M}{\sqrt{\ell_{\rm
              Planck}^2 k_i}}\right)}{\sqrt{\ell_{\rm
          Planck}^2 k_i}}\vert\psi,\tilde{g},\vec{k}\rangle_{\rm grav}, \\
\hat{B}_i \vert \psi,\tilde{g},\vec{k}\rangle_{\rm grav}&=&2 
  \frac{\left[\sqrt{\vert\Delta k_i+1\vert}-\sqrt{\vert\Delta k_{i}-1\vert}\right]^2}{\ell_{\rm Planck}^2}
    {\left(1+{\cal K}^2_{\varphi,i}
-\frac{2 G M}{\sqrt{\ell_{\rm
              Planck}^2 k_i}}\right)}k_i^{3/2} \ell_{\rm Planck}^3
    \vert \psi,\tilde{g},\vec{k}\rangle_{\rm grav},\\
\hat{C}_i\vert \psi,\tilde{g},\vec{k}\rangle_{\rm grav}&=&2 
  \frac{\left[\sqrt{\vert\Delta k_i+1\vert}-\sqrt{\vert\Delta k_{i}-1\vert}\right]^2}{\ell_{\rm Planck}^2}
\sqrt{1+{\cal K}^2_{\varphi,i}
-\frac{2 G M}{\sqrt{\ell_{\rm
              Planck}^2 k_i}}}
{\cal K}_{\varphi,i}\sqrt{\ell_{\rm
          Planck}^2 k_i}\vert\psi,\tilde{g},\vec{k}\rangle_{\rm grav}.
\end{eqnarray}
Notice that all dependence on $\Delta x$ has canceled out. Strictly
speaking, these expressions are valid for the action of the operators
on states of the form $\vert \tilde{g},\vec{k},M\rangle$. Acting on
states $\vert \psi,\tilde{g},\vec{k}\rangle_{\rm grav}$ these
expressions are only true at leading order in the dispersion of the
mass $\sigma$. From now on all expressions should be interpreted as
leading order in the dispersion of the mass.  We also have dropped the
subindex $0$ from $M_0$ for simplicity.

We will work in the exterior region where we can choose ${\cal K}_\varphi=0$
without incurring coordinate singularities, so the above expressions
simplify a bit, in particular $\hat{C}_i=0$ and,
\begin{eqnarray}
  \hat{A}_i \vert \psi,\tilde{g},\vec{k}\rangle_{\rm grav} &=& \frac{1}{2}
  \frac{1}{{\left(E^x\right)'_i}}
    \frac{\left(1
-\frac{2 G M}{\sqrt{\ell_{\rm
              Planck}^2 k_i}}\right)}{\sqrt{\ell_{\rm
          Planck}^2 k_i}}\vert\psi,\tilde{g},\vec{k}\rangle_{\rm grav}, \\
\hat{B}_i \vert \psi,\tilde{g},\vec{k}\rangle_{\rm grav}&=&2 
  \frac{1}{{\left(E^x\right)'_i}}
    {\left(1
-\frac{2 G M}{\sqrt{\ell_{\rm
              Planck}^2 k_i}}\right)}k_i^{3/2} \ell_{\rm Planck}^3
    \vert \psi,\tilde{g},\vec{k}\rangle_{\rm grav},
\end{eqnarray}
where,
\begin{equation}
  \frac{1}{\left(E^x\right)'_i}\equiv 
\frac{1}{{\ell_{\rm Planck}^2}}\left[\sqrt{\vert\Delta k_i+1\vert}-\sqrt{\vert\Delta k_{i}-1\vert}\right]^2.
\end{equation}

Taking the expectation value of the
Hamiltonian in the normalized state considered leads to an expression (recall
that in the exterior ${\cal K}_\varphi=0$) for the quantum Hamiltonian of
matter given by,
\begin{equation}\label{18}
  H= \sum_j A_j P_{\phi,j}^2+B_j\left(\phi_{j+1}-\phi_j\right)^2,
\end{equation}
with $A_j$ and $B_j$ the eigenvalues obtained above.

\section{The equations of motion}

We will now consider the equations of motion for the scalar field
stemming from the classical version of the Hamiltonian derived in the
previous section.  The associated classical equation 
of motion for the field is,
\begin{equation}
  \dot{\phi}_i= \left\{ \phi_i, H\right\}=2 A_i P_{\phi,i}.
\end{equation}
with $H$ given by (\ref{18}) with $P_{\phi_j}$ and $\phi_j$ classical
fields. One can introduce the associated Lagrangian,
\begin{equation}
  L = \sum_j P_{\phi,j}\dot{\phi}_j -H = \sum_j
  \frac{\dot{\phi}_j^2}{2 A_j}-B_j \left(\phi_{j+1}-\phi_j\right)^2,
\end{equation}
which can be recognized as a discrete version of,
\begin{equation}
  L = \int \sqrt{-g} \left(g^{00} \left(\partial_0 \phi\right)^2 +
g^{xx} \left(\partial_x \phi\right)^2\right)dx.
\end{equation}
To see this, let us consider its discretization,
\begin{equation}
  L = \sum_j \sqrt{-g(x_j)} \left(g^{00}(x_j) \left(\partial_0 \phi(x_j)\right)^2 +
g^{xx}(x_j) \frac{\left(\phi(x_{j+1})-\phi(x_j)\right)^2}{\Delta^2}\right) \Delta.\label{22}
\end{equation}
And from it we can read off the components of the discrete metric,
\begin{eqnarray}\label{28}
\sqrt{-g(x_j)}g^{00}(x_j)= \frac{1}{2 A_j \Delta},\\
\sqrt{-g(x_j)}g^{xx}(x_j)= B_j \Delta.  \label{29}
\end{eqnarray}
We should recall that these expressions are valid in the exterior only
($j>0$). The resulting expressions for
the covariant form of the
metric are,
\begin{eqnarray}
  g_{00}(x_j) &=& -1+\frac{2 G M}{\sqrt{\ell_{\rm Planck}^2 k_j}}, \\
  g_{xx}(x_j) &=& \frac{1}{1-\frac{2 G M}{\sqrt{\ell_{\rm Planck}^2 k_j}}}, 
\end{eqnarray}
which can be seen to coincide with the expectation value of the
metric, viewed as a parameterized Dirac observable, on the
gravitational states considered using the results of section III,
recalling that $g_{xx}=(E^\varphi)^2/E^x$ and $g_{00}=-(N^{\rm orig})^2$. We
recognize the usual form of the Schwarzschild metric. To obtain the
above expressions we used that $\left(E^x\right)'_i= 2 x_i \Delta$ in
the spin network chosen.

The equation of motion for the field becomes a spatially discretized
version of the Klein--Gordon equation in a curved space-time,
\begin{equation}
  \left({\sqrt{-g} g^{ab}} \phi_{,a}\right)_{,b}=0.
\end{equation}
For the kind of quantum states here considered for the gravitational
field one recovers what would be a lattice version of the equations of
a quantum field in a background space time. If one considered states
that involve superpositions the situation is more involved.

\section{The quantum vacua on a quantum background}

\subsection{Introduction}

The construction of quantum vacua on curved space-times is carried out
by considering modes that solve the wave equation on the curved
space-time and in terms of them constructing the creation and
annihilation operators \cite{vacua}. These constructions are slicing
dependent. Accordingly, different vacua have been defined in the
literature through the choice of different slicings. When one computes
physical quantities, like the expectation value of the stress energy
tensor in the vacuum, one may encounter singularities at various
points of space-time, particularly at the horizons. The Boulware
vacuum is associated with the tortoise radial coordinate and the time
of an observer at infinity. It leads to singularities of the physical
quantities at the past and future horizons. The Hartle--Hawking vacuum
is based on the Kruskal coordinates and does not lead to singularities
in the physical quantities anywhere. The Unruh vacuum is associated to
a foliation in which the Cauchy surfaces approach the past horizon and
the past null infinity asymptotically and leads to singularities in
the physical quantities at the past horizon. The Hawking radiation can
be computed by comparing the modes of the Unruh and Boulware vacua and
computing the expectation value of the 
number operator associated with one vacuum on the other. The
Hartle--Hawking vacuum has ingoing and outgoing modes and is therefore
associated with a black hole with incoming radiation in addition to
the Hawking radiation. 

We will start with a discussion of the coordinate systems used in the
definition of the vacua.

\subsection{Coordinate systems}

The first coordinate change we want to consider is to pass to a
tortoise radial coordinate. In the discrete case this corresponds to, 
\begin{equation}
  x^*_j=x_j+2 G M \ln\left(\frac{x_j}{2 G M}-1\right).
\end{equation}
The inverse transformation can be cast in terms of the Lambert
function $W$,
\begin{equation}\label{34}
  x_j=2 G M
\left(W\left(\exp\left(\frac{x^*_j-2 G M}{2 G M}\right)\right)+1\right).\label{lambert}
\end{equation}
Choosing a function $z(x(x^*))$ with $x(x^*)$ given by (\ref{34}),
reworking expressions (\ref{28},\ref{29}) leads to a metric,
\begin{eqnarray}
  g_{00}(x_j) &=& -1+\frac{2 G M}{\sqrt{\ell_{\rm Planck}^2 k_j}}, \\
  g_{xx}(x_j) &=& 1-\frac{2 G M}{\sqrt{\ell_{\rm Planck}^2 k_j}}. 
\end{eqnarray}

We will also need null coordinates,
\begin{eqnarray}
  u_i&=&t-x^*_i, \qquad x^*_i=\frac{v_i-u_i}{2},\\
  v_i&=&t+x^*_i, \qquad t=\frac{u_i+v_i}{2},
\end{eqnarray}
in terms of which the metric has the non-vanishing component, 
\begin{equation}
  g_{uv}(u_j,v_j)=-\frac{1}{2}\left(1-\frac{2 G M}{\sqrt{\ell^2_{\rm Planck} k_j}}\right),
\end{equation}
where $\ell_{\rm Planck}^2 k_j^2=x^2_j$ with the right hand side given
as function of $u_j,v_j$ via the Lambert function.

We will also need the Kruskal coordinates,
\begin{eqnarray}
  U_i&=& -\exp\left(\frac{u_i}{4 G M}\right),\\
  V_i&=& \exp\left(\frac{v_i}{4 G M}\right),
\end{eqnarray}
in terms of which the metric has a non-vanishing component,
\begin{equation}
  g_{UV}\left(U_j,V_j\right)= -4\frac{\left(2 G M\right)^3}{\sqrt{\ell_{\rm
      Planck}^2 k_j}}\exp\left(-\frac{\sqrt{\ell^2_{\rm Planck} k_j}}
{2 G M}\right),
\end{equation}
and we also have that 
\begin{equation}
  U_j V_j = -\exp\left(-\frac{x_j}{2 G M}\right)\left(\frac{x_j}{2 G M}-1\right).\label{76}
\end{equation}
These coordinate changes can all be done at the level of
(\ref{28},\ref{29}) defining appropriate functions $z(x)$ (in general
$z(x,t)$). We just proceeded directly since the changes are simpler to
do that way, remembering that the ($t,r$) portion of the metric is
explicitly conformally flat, and the changes preserve that nature.

A final comment about coordinates is that we will be studying
behaviors at scri. Strictly speaking, one cannot study null infinity
with our canonical framework based on a spatial spin network with a
finite number of points, so we will be really making statements about
behaviors at null surfaces far away from the black hole and into the
past or future, rather than scri itself. 

Notice that these coordinate manipulations just involve a relabeling
of the spin network vertices, not a change in their physical
distance. When the relabellings involve time, one is considering the
fields at different Schwarzschild times.

\subsection{Boulware vacuum}

The Boulware vacuum is constructed on a foliation determined by
coordinates $t,x^*$. In such a foliation, the Lagrangian we identified
for the discrete theory (\ref{22}) becomes, 
\begin{equation}
  L=
-\sum_j  \left(\left(\partial_0 \phi(x^*_j,t)\right)^2 -
\frac{\left(\phi(x^*_{j+1},t)-\phi(x^*_j,t)\right)^2}{\Delta_j^2}\right)
\Delta_j \label{77}
\end{equation}
where $\Delta_j$ is the separation between vertices in the tortoise
coordinate, which is non-uniform. Here we have made the simplifying
assumption of going to the $1+1$ dimensional case, ignoring the
centrifugal and gravitational potential that appear in $3+1$
dimensions with spherical symmetry on a curved background. For the
calculation of the Hawking radiation one needs the modes close to the
horizon and infinity only, and there the potential vanishes and the
discussion of the modes in both cases is equivalent. The equation of
motion that follows from this Lagrangian is, after some rescalings,
\begin{equation}
  \partial_0^2
  \phi(x^*_j,t)-\left[\frac{\phi(x^*_{j+1},t)-\phi(x^*_j,t)}{\Delta_j^2}
-\frac{\phi(x^*_{j},t)-\phi(x^*_{j-1},t)}{\Delta_j\Delta_{j-1}}\right]=0.\label{78}
\end{equation}
The separation in the tortoise coordinate is given by,
\begin{equation}
  \Delta_j = \frac{dx^*}{dx}\biggr\vert_{x_j} \Delta= \frac{x_j
    \Delta}{x_j-2GM}=\Delta+\frac{2 G M}{j},
\end{equation}
where we took into account that $j\Delta=x_j-2GM$. Asymptotically far
from the black hole $\Delta_j\to \Delta$ and one therefore recovers a
very refined equally spaced lattice. That leads to an excellent
lattice approximation to the traditional solutions for the modes that
define the Boulware vacuum,
\begin{eqnarray}
  f_\omega&=&\frac{1}{\sqrt{2\pi\omega}}e^{-i\omega(t+x^*)},\qquad
  g_\omega=\frac{1}{\sqrt{2\pi\omega}}e^{-i\omega(t-x^*)},\\
  \phi&=&\int_0^\infty d\omega \left(b_\omega f_\omega+b^\dagger_\omega
    f^*_\omega +c_\omega g_\omega +c^\dagger_\omega g^*_\omega\right),
\end{eqnarray}
where $b,c$ are creation and annihilation operators. 
The main difference with respect to the usual treatment on a classical
background is that we are effectively on a lattice and therefore the
dispersion relation is modified, taking the usual lattice form,
\begin{equation}
  \omega_n=\sqrt{\frac{2 -2 \cos\left(k_n \Delta \right)}{\Delta^2}},
\end{equation}
with $k_n=2 \pi n/((V-i_H)\Delta)$. The above expressions for the
modes are therefore good approximations for sub-Planckian frequencies,
where the dispersion relation is approximately linear. 

The appearance of trans-Planckian modes could be problematic close to
the horizon, since modes coming in from infinity get blueshifted. The
expected result of the introduction of a lattice is that the modes
with wavelengths shorter than the lattice get suppressed. To see this,
we would like to study the behavior of the modes close to the horizon.
We seek separable solutions $\phi(x^*_j,t)=A(t) B(x^*_j)$ to
(\ref{78}). We get the equations,
\begin{eqnarray}
  \ddot{A}(t)&=&-\omega^2 A(t),\\
-\omega^2&=& \frac{B\left(x^*_{j+1}\right)-B\left(x^*_{j}\right)}{
B\left(x^*_{j}\right)\Delta_j^2}-
\frac{B\left(x^*_{j}\right)-B\left(x^*_{j-1}\right)}{
B\left(x^*_{j}\right)\Delta_j\Delta_{j-1}}.
\end{eqnarray}
The solution to the first equation is immediate. The second can be
solved numerically, but close to the horizon $j\ll 2 GM/\Delta$ and
the equation takes the form,
\begin{equation}
B\left(x^*_{j+1}\right)=-\Delta_j^2\omega^2 B\left(x^*_{j}\right)  
+B\left(x^*_{j}\right)\left(\frac{2j-1}{j}\right)-
B\left(x^*_{j-1}\right)\left(\frac{j-1}{j}\right).
\end{equation}
Let us consider the case of waves with the shortest
wavelength permitted by the lattice and the case of long wavelengths. 
The maximum frequency allowed by the lattice is $\omega=2/\Delta$. 
In that case the first term on the right hand side dominates. Setting
a boundary condition $B(x^*_1)=B_0={\rm constant}$, the equation becomes,
for the two next points,
\begin{equation}
  B\left(x^*_3\right)=-\left(\frac{2 G M}{\Delta}\right)^2 
B\left(x^*_2\right)-\frac{1}{2}B_0,
\end{equation}
so we see that the values for successive lattice points alternate in
sign and increase rapidly as one moves away from the horizon. That
means that ingoing, these modes are heavily suppressed and one faces
no trans-Planckian problem in the vicinity of the horizon. For
sub-Planckian frequencies, the modes behave as in the continuum
starting at scri and then get modified close to the horizon. The
modification starts closer to the horizon for smaller $\Delta$'s
and/or smaller frequencies. There is a maximum frequency that
corresponds to $\omega=2/\Delta$.  On the other hand, for large
wavelengths $\omega=2 \pi/\lambda$ with $\lambda\gg 2 GM$, one can see
that the solution agrees very well with the continuum solution with
small deviations, as one expects from a lattice approximation with
small lattice spacing compared to the wavelength.

The fact that the high frequency modes suffer significant changes in the
region close to the horizon(s)  has consequences for physical
quantities.  In particular, physical quantities
computed with the quantum states stop being singular there. In fact
they are non-singular everywhere. The singularities in the usual
quantum field theory in curved space time framework
manifest themselves in the Feynman propagators and the expectation
values of the energy momentum tensor . The propagators have singular
derivatives on the horizon(s) and as most physical properties of the
matter field ---like the energy--- involve derivatives, their physics
becomes singular there. This is due to the fact that in the continuum
the radiation never enters the horizon in these coordinates and there
are trans-Planckian modes due to the infinite blueshift of the
radiation at the horizon. This is the origin of the singularities
associated with the Boulware vacuum at the horizon in the traditional
treatment. In our treatment there are no trans-Planckian modes, the
dispersion relation is modified in a sub-luminal way (it does not
affect the horizon structure) and there are no singularities. Also
the radiation reaches the horizon in a finite time, since the last
point before the horizon is reached in a finite time and from the
horizon to the interior one can choose a non-vanishing $K_\varphi$ and
transition seamlessly without coordinate singularities.  The
appearance of a fundamental discrete lattice, in addition to
eliminating the singularities in the physical quantities, also
eliminates the ambiguities present in the continuum in the
regularization of quantities, like for instance in the energy-momentum
tensor. This has been a major obstacle to performing back reaction
calculations, which would not be present in the current approach.

\subsection{The Hartle--Hawking vacuum}

As in the usual treatment, the modes are perfectly well behaved near the
horizon and they are not infinitely blue shifted as there is no
coordinate singularity at the horizon.  In this case the
uniform spacing in the radial coordinate translates itself into a
spacing that grows when one approaches scri. The Hartle--Hawking modes
are formulated in terms of a chart $T,R$ given in terms of the Kruskal
coordinates by $R=V-U$ and $T=V+U$. In a surface $T=T_0={\rm
  constant}$, we have that
\begin{equation}
  R^2-T_0^2=\exp\left(\frac{r}{2GM}\right)\left(\frac{r}{2GM}-1\right),
\end{equation}
which implies
\begin{equation}
  R=\sqrt{4 \exp\left(\frac{r}{2GM}\right)\left(\frac{r}{2GM}-1\right)+T_0^2},
\end{equation}
and therefore,
\begin{equation}
  \Delta R = \frac{\partial R}{\partial r} \Delta =\frac{\sqrt{2 r}
    \exp\left(\frac{r}{4 GM}\right)}{\left(2 GM\right)^{3/2}}\Delta,
\end{equation}
which grows exponentially with $r$, no matter how small one takes
$\Delta>\ell_{\rm Planck}/(2r)$. As a consequence, some modes of the
continuum cannot be reproduced. However, such modes are not physically
relevant, since they would imply trans-Planckian frequencies at scri
and there is no good motivation to consider such frequencies there.

\subsection{The Unruh vacuum}

Finally, for the Unruh vacuum one chooses a foliation with coordinates
$\tilde{x}^U=\ln(V/2)-U/2$, $\tilde{t}^{U}=\ln(V/2)+U/2$. Here we will
make a slightly different choice of congruence in order to enhance the
behavior at the past horizon,
\begin{eqnarray}
  \tilde{x}^U&=&\frac{V+1}{V}\ln(V/2)-U/2,\\
\tilde{t}^{U}&=&\frac{V+1}{V}\ln(V/2)+U/2.
\end{eqnarray}
The additional factor in front of the logarithm has the role of
keeping the $\tilde{t}^U={\rm constant}$ surfaces closer to the past
horizon when one discretizes. Otherwise one would, in the first
discrete vertex away from the horizon allowed by the quantization of
area condition, be already very far away from it, and the region close
to the horizon would contain very few vertices, spoiling the continuum
approximation. The extra factor does not alter the global behavior of
the slicing, in particular near the future horizon one has incoming
modes that come from scri- with positive frequency in terms of the
standard Schwarzschild time $t$. 

The metric takes the following form,
\begin{eqnarray}
  g_{00}^U&=&-\frac{32 \left(G M\right)^3}{r}
  \frac{V^2}{V+1-\ln\frac{V}{2}}\exp\left({-\frac{r}{2 G M}}\right),\\
  g_{xx}^U&=& -g_{00}^U,
\end{eqnarray}
where $r=r(U,V)$ given by inverting (\ref{76}). The extra factor
involving $V$ tends exponentially to infinity at scri+ and compensates
the exponentially decreasing term in $r$, the metric does not go to
zero at scri+, as was the case in the slicing of the Hartle--Hawking
vacuum. This will lead to reproducing the continuum modes at
scri+. 
The separation between vertices tends to $\Delta$ at scri+, $\delta
x^U\to2 \Delta/(2 G M)$  when $r\to \infty$. Since the metric is
conformal in these coordinates, the modes coincide with those of flat
space-time in the $\tilde{t}^U, \tilde{x}^U$ coordinates. When written
in terms of the Kruskal coordinates $U,V$, the modes are $\exp(i\omega
U)/\sqrt{2\pi \omega}$ and $\exp(2\ln(V/2))/\sqrt{2\pi \omega}$. 

Since we have no problems at infinity, we will study the behavior at
the past horizon. For this, we consider a Cauchy surface for the
outgoing modes $\tilde{t}^U=t_0\ll 0$. For $V\sim 1$, $\tilde{t}^U$
starts to quickly depart from the past horizon. Since  for large $V$ we
have that $V=2\exp((\tilde{x}^U+\tilde{t}^U)/2)$ and
$U=\tilde{t}^U-\tilde{x}^U$, we have that $V\sim 1$ occurs for
$\tilde{x}^U\sim-t_0$. The variable $\tilde{x}^U$ ranges from
$\tilde{x}^U=t_0$ in which case $U=0$ and one is at the future
horizon, all the way to $\tilde{x}^U=-t_0$ where one is at the past
horizon. To understand the behavior at the past horizon we have to
look at very negative values of $t_0$. An important property of the
Unruh coordinates is that the potential term in the equation for the
modes at $\tilde{t}^U=t_0$ is only significantly different from zero
at $\tilde{x}^U=-\tilde{t}^U$, where we have that $V=1$ and the
initial surface quickly departs from the past horizon. In the limit
$\tilde{t}^U\to -\infty$ the complete horizon is covered by the
surface and the potential tends to zero. However, that limit
corresponds to $V\to0$ and one immediately sees the metric is singular
there. This is the origin of the singularities in physical quantities
of the Unruh vacuum at the past horizon. In our
case, the spin network never reaches the horizon so this problem is
avoided. 

To study the influence of the discrete structure in the behavior of
the solutions at $\tilde{t}^U=t_0$ we note that the vertices of the
spin network lie on curves $U_j V_j={\rm constant}$. We therefore
have,
\begin{equation}
  U_j V_j=-\exp\left(\frac{x_j}{2 G M}\right)
\left(\frac{x_j}{2 G M}-1\right),
\end{equation}
so at $\tilde{t}^U=t_0$ one has that,
\begin{equation}
\frac{  V_j(t)+1}{V_j(t)}\ln\left(V_j(t)\right)
-\frac{1}{V_j(t)}\exp\left(\frac{x_j}{2 G M}\right)
\left(\frac{x_j}{2 G M}-1\right)=t_0.
\end{equation}
From here one can read off the distribution of $V_j(t)$ and from
$U_jV_j={\rm constant}$ one can see that the points $U_j(t)$ are
approximately uniformly distributed between $U=0$ and $U=2t_0$, which
is the region in which $\tilde{t}^U=t_0$ approximates the past
horizon. As a consequence, the spacing between vertices of the spin
network  in terms of $U$ is given by  $\delta U=2 t_0 \Delta/(2 G
M)$. The spacing therefore is a function of time. To study the
propagation of modes, we consider a lattice with spacing that is time
dependent but spatially uniform. 

So we start from equation (\ref{78}) considering a time dependent
spacing (notice that such equation would not follow from the action
(\ref{77}), a complete treatment would have required re-deriving
(\ref{77}) with ${\cal K}_\varphi$ and $z$ functions of $x$ and
$t$. This in particular would have led to a complicated dependence
of the lattice spacing with space and time, and we could not find a
way to treat the situation analytically, so we are using here an
approximation in which the lattice spacing only depends on time), 
\begin{equation}
  \partial_0^2
  \phi(\tilde{x}^U_j,\tilde{t}^U)-\left[\frac{\phi(\tilde{x}^U_{j+1},\tilde{t}^U)-2\phi(\tilde{x}^U_j,\tilde{t}^U)+\phi(\tilde{x}^U_{j-1},\tilde{t}^U)}{a(\tilde{t}^U)^2}\right]=0,
\end{equation}
where in our case $a(\tilde{t}^U)=2 \tilde{t}^U \Delta/(2 G M)$. It is
worthwhile reminding that we are working on a lattice with a finite
number of points $N$ in the region close to the horizon and up to the
maximum of the centrifugal potential in order to characterize these
modes. So the total size of the lattice will change as the spacing
changes with time. We note that the above equation can be solved by
the ansatz,
\begin{equation}
  \phi(\tilde{x}^U_j,\tilde{t}^U)=f\left(\tilde{t}^U\right)\exp\left(i
    \tilde{x}^U_j \frac{2\pi n}{N
      a(t)}\right)=f\left(\tilde{t}^U\right)\exp\left(i \frac{2\pi n j}{N}\right),
\end{equation}
with $f\left(\tilde{t}^U\right)$ satisfying,
\begin{equation}
    a\left(\tilde{t}^U\right)^2 \frac{\ddot{f}}{f}= 2 \cos\left(\frac{2 \pi n}{N}\right)-2=-K^2_n,\label{98}
 \end{equation}
which with the definition of $a(\tilde{t}^U)$ yields,
\begin{equation}
  \ddot{f} +\frac{L^2_n}{t^2}f=0
\end{equation}
with $L^2_n={(G M K_n )^2}/{\Delta}^2$ (and for a stellar sized
black hole $L_n\sim 10^{67}$). Solving the equation we get,
\begin{equation}
  f\left(\tilde{t}^U\right)=A \exp\left(\frac{1}{2} \pm \frac{i}{2} \sqrt{4 L^2_n-1}\right)\ln\left(-\tilde{t}^U\right).
\end{equation}

To understand this solution we set ourselves in an asymptotic past
time $\tilde{t}^U_1$ so it takes a large negative value and we study
how the solution evolves for a period of time $\tau$ much smaller than
$\tilde{t}^U_1$, 
\begin{equation}
  f\left(\tilde{t}^U_1+\tau\right) = A(\tilde{t}^U_1) \exp\left(\pm i \frac{L_n}{t_1}\tau+\frac{\tau}{2\tilde{t}^U_1}\right),
\end{equation}
with $A(t_1)$ a constant. So apart from a term that changes the
amplitude, it has an oscillation with a frequency that varies with
time. For $\tilde{t}^U_1\to -\infty$ the frequency is zero, as the
metric vanishes on the past horizon. Computing the physical separation
between points in the lattice as one goes near the horizon, it tends
to the Planck length (choosing the most refined spin network subject
to the condition of quantization of area).  There exists a cutoff
frequency that is coordinate independent and trans-Planckian modes are
suppressed. The presence of the discrete structure modifies the
dispersion relations. One can work them out explicitly expanding 
for small momenta equation (\ref{98}) the quantity 
$2\pi n/N$ plays the role of
momentum in the lattice, whereas $2\pi n/(N a(\tilde{t}^U))$ is the
physical momentum).

\section{Hawking radiation}

In order to compute the Hawking radiation we will use the technique of
computing the two point functions and from them compute the
expectation value of the number operator \cite{valencia}. We will not 
strictly carry out the discrete calculations associated with the spin
network background. We will sketch the usual
calculations on a classical space time and point out at key instances
what differs in the discrete case. 

The main ingredient is the calculation of the Bogoliubov coefficients,
which connect the modes of the ``in'' and ``out'' states,
corresponding respectively to the state of the field at the past
horizon and at future scri, and are defined as,
\begin{equation}
\beta_{i_1,k}=  \left(u_{i_1}^{\rm out},{u_{k}^{\rm in}}^*\right),
\end{equation}
and we introduce the Klein--Gordon inner products among modes $u_{k}$
defined by
\begin{eqnarray}
  \left(u_{i_1}^{\rm out},{u_{k}^{\rm
        in}}^*\right)&=&\int_{{\cal I}+} d\Sigma^\mu u^{\rm
    out}_{i_1}\left(x\right)\stackrel{\leftrightarrow}{\partial}_\mu 
u^{\rm
    in}_{k}\left(x\right),\\
\left({u_{i_2}^{\rm out}}^*,{u_{k}^{\rm in}}\right)&=&
\int_{{\cal I}+} d\Sigma^\mu u^{\rm
    out}_{i_2}\left(x\right)\stackrel{\leftrightarrow}{\partial}_\mu 
u^{\rm
    in}_{k}\left(x\right),
\end{eqnarray}
where the integrals are calculated at scri+ ${\cal I}$. Since we are
in the spherical case we will consider the modes, 
\begin{eqnarray}
  u^{\rm in}_{\omega,\ell,m} &=&\frac{1}{\sqrt{4 \pi
    \omega}}\frac{\exp\left(-i\omega U\right)}{r} Y_\ell^m\left(\theta,\varphi\right)\\
  u^{\rm out}_{\omega,\ell,m} &=&\frac{1}{\sqrt{4 \pi
    \omega}}\frac{\exp\left(-i\omega u\right)}{r} Y_\ell^m\left(\theta,\varphi\right)
\end{eqnarray}
where $\omega,\ell,m$ play the role of $k$\footnote{To simplify
  expressions, in this section we switched conventions and are using a
  variable $U$ that has dimensions of length, as does $u$. So this $U$
  is $4GM$ times the previously defined one.}.  Given the relation
between the creation and annihilation operators out and in given by
the Bogoliubov coefficients,
\begin{equation}
  a^{\rm out}_{i_1}= \sum_k \left(\beta_{i_1,k} a^{\rm in}_k +
\beta^*_{i_1,k} {a^{\rm in}}^\dagger_k\right), 
\end{equation}
we can compute the expectation value of the number operator associated
with the out modes in the in states, which is non-vanishing,
\begin{equation}
  \langle {\rm in}\vert N^{\rm out}_{i_1,i_2}\vert {\rm in}\rangle = 
\sum_k \beta_{i_1,k}\beta^*_{i_2,k}=-\sum_k \left(u_{i_1}^{\rm out},{u_{k}^{\rm in}}^*\right)\left({u_{i_2}^{\rm out}}^*,{u_{k}^{\rm in}}\right),
\end{equation}

We note that the sum in $k$ that appears is the two point function,
\begin{eqnarray}
  \langle {\rm in}\vert \phi\left(x_1\right)
\phi\left(x_2\right)\vert {\rm in}\rangle &=& \sum_k u^{\rm
  in}_k\left(x_1\right)
{u^{\rm in}}^*_k\left(x_2\right)\\
&=&\int_0^\infty d\omega \sum_{\ell,m} \frac{\exp\left(-i\omega
    U_1\right)}{\sqrt{4\pi\omega}}\frac{Y_\ell^m\left(\theta_1,\varphi_1\right)}{r_1}
\frac{\exp\left(-i\omega U_2\right)}{\sqrt{4\pi\omega}}\frac{{Y_\ell^m}^*\left(\theta_2,\varphi_2\right)}{r_2}.
\end{eqnarray}
So we can rewrite the expectation value of the number operator as,
\begin{eqnarray}
&&  \langle {\rm in}\vert N_{i_1,i_2}^{\rm out}\vert {\rm in}\rangle = 
\int_0^\infty r_1^2 dU_1 d\Omega_1 r_2^2 dU_2 d\Omega_2
\times\nonumber\\
&\times& \frac{\exp\left(-i\omega_1 u(U_1)\right)}{\sqrt{4 \pi \omega_1}}
\frac{Y_{\ell_1}^{m_1}\left(\theta_1,\varphi_1\right)}{r_1}
\frac{\exp\left(-i\omega_2 u(U_2)\right)}{\sqrt{4 \pi \omega_2}}
\frac{Y_{\ell_2}^{m_2}\left(\theta_2,\varphi_2\right)}{r_2}
\partial_{U_1}\partial_{U_2} \langle {\rm in}\vert
\phi\left(x_1\right)
\phi\left(x_2\right)\vert {\rm in}\rangle t_\ell\left(\omega_1\right)
t^*_\ell\left(\omega_2\right),
\end{eqnarray}
where we have integrated by parts in $U_1,U_2$.  The factors
$t_\ell(\omega)$ are the ``transmission coefficients'' given by the fact that
one has a potential in the wave equation and therefore not the
entirety of the modes get transmitted to infinity.
The angular integrals can be computed to give $4
\delta_{\ell_1,\ell_2} \delta_{m_1,m_2}$. In the case of quantum field
theory on a classical space-time the calculation continues as follow,
\begin{equation}
\partial_{U_1}\partial_{U_2} \langle {\rm in}\vert
\phi\left(x_1\right)
\phi\left(x_2\right)\vert {\rm in}\rangle =
  \partial_{U_1}\partial_{U_2}\int_0^\infty d\omega
\frac{\exp\left(-i\omega\left(U_1-U_2\right)\right)}{4 \pi \omega}=
\int_0^\infty d \omega
\frac{\omega}{4\pi}\exp\left(-i\omega\left(U_1-U_2\right)\right).
\end{equation}
Although the integral is immediate, its expression has a problem at
$\omega\to\infty$, where it oscillates. To compute it in the
continuum, 
one introduces
a small purely imaginary term $i \epsilon$, that ensures an
exponential falloff for $\omega\to\infty$ and can be removed by taking
the limit $\epsilon\to0$,
\begin{equation}
\lim_{\epsilon\to 0} i \partial_{U_1}\int_0^\infty
\frac{d\omega}{4\pi}\exp\left(-i\omega\left(U_1-U_2-i\epsilon\right)\right)
=\lim_{\epsilon\to 0}\frac{1}{4\pi}\frac{1}{\left(U_1-U_2-i\epsilon\right)^2}.
\end{equation}
In our quantum treatment, due to the discreteness introduced by the
spin network one gets similar expressions but there exists a maximum
$\omega=4 \pi G M/ \ell_{\rm Planck}^2$. One cannot take the limit
$\epsilon\to 0$ and the parameter must take a small finite value. We
need $\exp(-\omega \epsilon)$ to be small, so that leads to a value
for $\epsilon\sim \ell_{\rm Planck}$ to avoid getting frequencies
where the dispersion relation differs from the continuum. This may
sound ad-hoc presented this way, but there is a good motivation for
it. We are considering only the leading contributions in the
dispersion of the mass. In reality one will have states that are
superpositions of states with different values of the mass. The
natural value for the mass uncertainty is given by the Planck
scale. This introduces a fuzziness in the lattice, that leads to the
elimination of frequencies higher than the Planck one where the
dispersion relation differs from the continuum one.

This leads to an expression for the expectation
value of the number operator is,
\begin{equation}
  \langle {\rm in}\vert N^{\rm out}_{i_1,i_2}\vert {\rm in}\rangle = A
  \int_{I+} dU_1 dU_2 \frac{\exp\left(-i\omega_1 u_1(U_1)+i\omega_2
      u_2(U_2)\right)}
{\left(U_1-U_2-i\epsilon\right)^2},
\end{equation}
with
\begin{equation}
  A=\frac{t_{\ell_1}\left(\omega_1\right)t^*_{\ell_2}\left(\omega_2\right)}
{4\pi^2 \sqrt{\omega_1\omega_2}}\delta_{\ell_1,\ell_2}\delta_{m_1,m_2}.
\end{equation}

We now perform a change of variables,
\begin{eqnarray}
  U_1&=&-4 G M \exp\left(\frac{u_M+z}{4 G M}\right),\\
  U_2&=&-4 G M \exp\left(\frac{u_M+z}{4 G M}\right),
\end{eqnarray}
where $u_M=u_1+u_2$ and $z=u_2-u_1$,
and the calculation reduced to an integral in $z$ that can be computed,
\begin{equation}
    \langle {\rm in}\vert N^{\rm out}_{i_1,i_2}\vert {\rm in}\rangle =
\frac{t_{\ell_1}(\omega_1)t_{\ell_2}^*(\omega_2)\delta(\omega_1-\omega_2)}{2
  \pi \sqrt{\omega_1\omega_2}} \int_{-\infty}^\infty
dz\exp\left(-i\frac{\left(\omega_1-\omega_2\right)}{2}z\right)
\left(\frac{\kappa}{2}\right)^2 \frac{\delta_{\ell_1,\ell_2}\delta_{m_1,m_2}}{\sinh^2\left(\frac{\kappa}{2}\left(z-i\epsilon\right)\right)},
\end{equation}
where $\kappa=1/(4 G M)$. 
Performing the integral in $z$ yields the Hawking formula,
\begin{equation}
    \langle {\rm in}\vert N^{\rm out}_{i_1,i_2}\vert {\rm in}\rangle =
\frac{\vert t_\ell(\omega)\vert^2}{\exp\left(2 \pi \omega/k\right)-1}.
\end{equation}

In the quantum background case, the spin network introduces a
cutoff in length  $\ell^2_{\rm Planck}/(4 G M)$, but as we mentioned,
we are considering wavelengths much larger than such cutoff to avoid
running into the modified dispersion relation, this implies 
a  cutoff of the order $\ell_{\rm
  Planck}$ in the $u$ variable. One therefore has
\begin{equation}
  \left(u_1-u_2\right)^2 \ge {\ell_{\rm Planck}^2}
\end{equation}
which is the same as that of \cite{valencia}. Notice that the cutoff
emerging in the $u$ variable is important since otherwise Lorentz
invariance would be violated. This leads to the same formula found in
\cite{valencia},
\begin{equation}
    \langle {\rm in}\vert N^{\rm out}_{i_1,i_2}\vert {\rm in}\rangle =
\frac{\vert t_\ell(\omega)\vert^2}{\exp\left(2 \pi \omega/\kappa\right)-1}
-\frac{\kappa^2 \ell_{\rm Planck}}{96 \pi^3 \omega}
\end{equation}
with the second term much smaller than the first one at least for
black holes with Schwarzschild radius bigger than the Planck length
and typical frequencies.  A more complete study considering
non-vanishing dispersion of the mass of the background state, which
will lead to a formula valid for all frequencies, with significant
modifications at high frequencies, will be presented in a forthcoming
publication.

Let us recall that in the canonical framework we are using, spin
networks live on Cauchy surfaces, so strictly speaking the treatment
we have done in this section takes place on a null surface far into
the future and far from the black hole, but not technically on scri,
which cannot be reached by the spin networks we consider.

\section{Conclusions}

We have studied the quantization of a scalar field on a quantum space
time that approximates well the geometry of a Schwarzschild black
hole. The treatment reproduces the results of quantum field theory on
a classical space-time well, with some interesting differences. The
presence of a discrete structure for the space-time eliminates the
divergences associated with the Boulware and Unruh vacua 
arising from the trans-Planckian
modes and only slight modifications for the Hartle--Hawking vacuum.  All
the different vacua's modes change considerably on the horizons where all the
singularities present in the usual analysis disappear. 

We have carried out the analysis for a given spin network, but it is
valid and can be extended without significant changes (except the one
we will mention next) for generic refinements of the given spin
network that include more vertices such that $k_n$ grows monotonically
with $n$.  The main difference with the behavior of the vacua in the
continuum is the loss of local Lorentz invariance, that in the state
we considered, would lead to corrections in the propagator in the
asymptotic region of scale
$\Delta x^2 p^4$. It is worthwhile pointing out that there exist
states in the physical space of states where the corrections are much
smaller than the Planck scale and therefore it is possible that the
argument of Collins {\em et al.} \cite{collins} that implies that
interactions in the perturbative treatment will lead to unacceptably
large corrections to observable quantities may not be applicable. As
it was emphasized in \cite{lorentz,polcho}, the Collins {\em et al.}
argument requires that the corrections be large at the Planck
scale. To see that they are not large we note that $r^2$ is $\ell_{\rm
  Planck}^2 k$ with $k$ an integer. One can have discretizations where
$k$ differ at most in one unit. In that case the radial distance goes
from $r=\ell_{\rm Planck}\sqrt{k}$ to $\ell_{\rm
  Planck}\sqrt{k+1}$. Since $\sqrt{k} =r/\ell_{\rm Planck}$ we have
that $\Delta r=\ell_{\rm Planck}^2/r<\ell_{\rm Planck}^2/(2GM)$. These
are corrections of order $\ell_{\rm Planck}^2$ and therefore are small
in the sense of the Collins {\em et al.} argument. It is premature to
try to draw conclusions about Lorentz invariance at this stage, a
proper study will require introducing interactions in the quantum
fields and study the radiative corrections.

The cutoff the type of discreteness here considered introduces is
similar in nature to the one considered by \cite{valencia} and leads 
to a similar calculation of the Hawking radiation, which does not suffer
significant modifications with respect to the continuum, at least for
large black holes and typical frequencies. 

Summarizing, we have shown that the midisuperspace formulation of loop
quantum gravity with spherical symmetry is able to reproduce many
features of standard analysis of quantum vacua in black hole
space-times in the limit in which one considers a quantum test field
living on a quantum space time. The discreteness of the quantum
space-time has implications for some of the vacua even in regions of
low curvature, in particular eliminating singularities. An interesting
task ahead is to study in detail the type of regularization that the
background introduces on the two point functions and to understand the
origin of the Hadamard conditions. This opens the possibility of
contemplating enhancing these computations with back reaction.  We
have only taken the first steps towards computing Hawking radiation in loop
quantum gravity. A more complete treatment, including superpositions
of the quantum spin network states and a more complete discussion of
the properties of the Green's functions of the theory, in particular
their Lorentz invariance will be pursued in further publications.

We wish to thank Iv\'an Agull\'o and Abhay Ashtekar for discussions.
This work was supported in part by grant NSF-PHY-1305000, funds of the
Hearne Institute for Theoretical Physics, CCT-LSU and Pedeciba.


\begin{references}


\bibitem{amps}
  A.~Almheiri, D.~Marolf, J.~Polchinski, D.~Stanford and J.~Sully,
  JHEP {\bf 1309}, 018 (2013)
  [arXiv:1304.6483 [hep-th]] and references therein.
\bibitem{akl}
  A.~Ashtekar, W.~Kaminski and J.~Lewandowski,
  Phys.\ Rev.\ D {\bf 79}, 064030 (2009)
  [arXiv:0901.0933 [gr-qc]].
\bibitem{dlt}
 A.~Dapor, J.~Lewandowski and Y.~Tavakoli,
  arXiv:1305.4513 [gr-qc].
\bibitem{sphericalprl} R.~Gambini and J.~Pullin,
  Phys.\  Rev.\  Lett.\  110, {\bf 211301} (2013)
  [arXiv:1302.5265 [gr-qc]].

\bibitem{rovelli} C. Rovelli, Phys. Rev. D42, 2638 (1990); Phys. Rev
  D43, 442 (1991). 



\bibitem{4models}  
R.~Gambini and R.~A.~Porto,
  Phys.\ Rev.\ D {\bf 63}, 105014 (2001)
  [gr-qc/0101057].

\bibitem{visser}
  C.~Barcelo, S.~Liberati and M.~Visser,
  Living Rev.\ Rel.\  {\bf 8}, 12 (2005)
  [Living Rev.\ Rel.\  {\bf 14}, 3 (2011)]
  [gr-qc/0505065].

\bibitem{thiemannqsd}
  T.~Thiemann,
  Class.\ Quant.\ Grav.\  {\bf 15}, 839 (1998)
  [gr-qc/9606089].
\bibitem{thiemanngiesel}
  K.~Giesel and T.~Thiemann,
  Class.\ Quant.\ Grav.\  {\bf 24}, 2499 (2007)
  [gr-qc/0607100].
\bibitem{asbole}
  A.~Ashtekar, M.~Bojowald and J.~Lewandowski,
  Adv.\ Theor.\ Math.\ Phys.\  {\bf 7}, 233 (2003)
  [gr-qc/0304074].
\bibitem{vacua}
N. Birrell,
P. Davies, ``Quantum fields in curved space'', Cambridge University
Press, Cambridge (1984);
R. M. Wald, ``General Relativity'', Chicago University Press, Chicago
(1984);  
  M.~Castagnino and R.~Ferraro,
  Phys.\ Rev.\ D {\bf 43}, 2610 (1991);
S. Carroll ``Spacetime and geometry: an introduction to
general relativity'', Addison--Wesley, New York (2003).
\bibitem{valencia}
  I.~Agullo, J.~Navarro-Salas, G.~J.~Olmo and L.~Parker,
  Phys.\ Rev.\ D {\bf 80}, 047503 (2009)
  [arXiv:0906.5315 [gr-qc]];   I.~Agullo, J.~Navarro-Salas, G.~J.~Olmo and L.~Parker,
  Phys.\ Rev.\ D {\bf 76}, 044018 (2007) [hep-th/0611355].

\bibitem{collins}  J.~Collins, A.~Perez, D.~Sudarsky, L.~Urrutia and H.~Vucetich,
  Phys.\ Rev.\ Lett.\  {\bf 93}, 191301 (2004)
  [gr-qc/0403053].
\bibitem{lorentz}
  R.~Gambini, S.~Rastgoo and J.~Pullin,
  Class.\ Quant.\ Grav.\  {\bf 28}, 155005 (2011)
  [arXiv:1106.1417 [gr-qc]]. 
\bibitem{polcho}
See also   J.~Polchinski,
  Class.\ Quant.\ Grav.\ {\bf 29}, 088001 (2012) [arXiv:1106.6346
  [gr-qc]].

\end{references}
\end{document}